\documentclass[12pt,preprint]{aastex}
\usepackage{epsfig}

\shortauthors{Chang, Konopelko, \& Cui} 
\shorttitle{PWN Associations with Unidentified TeV $\gamma- $Ray Sources}

\begin{document}

\title{Search for Pulsar Wind Nebula Association with Unidentified TeV $\gamma- $Ray Sources}

\author{Chulhoon~Chang, Alexander~Konopelko, and Wei~Cui}
\affil{Department of Physics, Purdue University, West Lafayette, IN 47907}
\altaffiltext{1}{Email: chang40@physics.purdue.edu, akonopel@purdue.edu, cui@physics.purdue.edu}

\begin{abstract}

Many of the recently discovered TeV $\gamma- $ray sources are associated with pulsar wind nebulae
(PWNe). In fact, they represent the most populous class of Galactic sources at TeV energies.
In addition, HESS has also discovered, in a survey of the Galactic plane, a population of TeV 
$\gamma- $ray sources that are still without definitive counterparts at longer wavelengths. 
For a number of these sources, a pulsar is an evident association, which is often located within 
an extended region of the TeV $\gamma- $ray emission. These particular HESS sources are promising
candidates for yet not resolved pulsar wind nebulae. Here we have undertaken a systematic search 
for X-ray counterparts of the sources, using the archival {\it Chandra} data, within the spatial 
bounds of the unidentified HESS sources. A number of X-ray sources have been detected in the 
{\it Chandra} fields. Two of them,  CXOU~J161729.3-505512 and CXOU~J170252.4-412848, are of a 
special interest because of their excellent positional coincidence with the pulsars 
PSR~J1617-5055 and PSR~J1702-4128, respectively. The first source is extended, with a bright 
core of $2.6\arcsec$ (FWHM) in radius but the emission can be seen up to roughly $20\arcsec$. 
The second one is much fainter and detected only with marginal 
significance ($4.6 \sigma$). It might also be slightly extended, although the situation 
is quite uncertain due to very limited statistics. The analysis of the archival {\it Chandra} 
data for a middle-aged pulsar (PSR~J1913+1011) does not reveal any statistically significant 
excess at and around the position of the pulsar or the center of gravity of its plausible
TeV $\gamma- $ray counterpart (HESS~J1912+101). We discuss the implications of the results.

\end{abstract}


\section{Introduction}

The charged particles accelerated in the vicinity of a pulsar flow out into the supernova (SN) 
ejecta and form a shock. The shock may further accelerate particles to relativistic speeds. 
These particles interact with the surrounding medium to produce a pulsar wind nebula (PWN),
which is often observable at radio and X-ray wavelengths. Around the youngest, most energetic 
pulsars, the radio emitting parts of these nebulae are rather amorphous, whereas the X-ray 
emitting regions can be highly structured \citep{GaenslerSlane2006}. The unprecedented spatial 
resolution of {\it Chandra} has made it possible to resolve these structures. The spatial 
extent and morphology of X-ray nebulae can vary from a toroidal (or arc-like) nebulae with 
perpendicular jets to the prominent tails behind the moving pulsars, indicative of bow 
shocks. The review of recent {\it Chandra} observational results on PWNe was given in 
\citet{KargaltsevPavlov2008}.

It was widely believed that PWNe could be sources of very high-energy (VHE) $\gamma- $ray 
emission. The emission may arise from inverse Compton scattering of low-energy photons by 
the relativistic electrons, while the X-ray emission may be associated with the synchrotron 
radiation from the same electrons. The best example is the Crab Nebula, which was established 
as a source of pulsed GeV $\gamma- $rays by EGRET \citep{Nolan1993} and of steady TeV $\gamma $ 
rays by ground-based \v{C}erenkov detectors \citep{Crab2006}. The TeV emission is thought to 
originate in its PWN \citep{deJagerHarding1992}. Recently, the HESS array of four imaging 
\v{C}erenkov telescopes have detected many TeV sources, a number of which are evidently 
associated with PWNe. 
Such associations usually rest on positional and morphological match of VHE $\gamma- $ray 
sources with known PWNe. It is worth noting that in many cases the pulsar and associated PWN 
are significantly offset from the center of the TeV source. The offset could be attributed to 
the interaction between the PWN and the SNR \citep{Blondinetal2001}.

A survey of the inner part of the Galactic Plane was performed with HESS between $\pm$30$^\circ$ 
in longitude and $\pm$3$^\circ$ in latitude relative to the Galactic Center. Fourteen previously
unknown, extended sources were detected with high significance \citep{aha2005,aha2006b}. 
Recently, the survey region has been extended to cover the 
longitude range out to +60$^\circ$ \citep{aha2008a}, leading to the discovery of five new TeV
sources. The HESS sources are likely Galactic in origin, given their concentration around the 
Galactic plane. The $\gamma- $ray flux of these sources varies from about 3\% to 25\% of that 
of the Crab Nebula. Most of them show rather hard energy spectra and are spatially extended, 
with the angular size of a few tenths of a degree. Some of the sources have fairly 
well-established counterparts at longer wavelengths, based exclusively on positional 
coincidence, but others have none at all. A number of proposals have been made on the nature of 
these unidentified TeV sources. At present, PWNe and shell-type SNRs are considered to be most 
probable, although other possibilities also exist. 

Observationally, a PWN often manifests itself as extended X-ray emission around a pulsar. The 
positional coincidence of such an object with an unidentified TeV source would strongly support 
the scenario that X-rays and TeV $\gamma- $rays are both powered by the PWN. For example, 
HESS~J1804-21, one of the brightest and most extended sources discovered in the HESS Galactic
plane survey, contains the Vela-like pulsar PSR~J1803-2137 at distance of about $3.8\, \rm kpc$ 
\citep{aha2005}. The pulsar has a spin-down age of about $16\, \rm kyr$ and a spin-down power 
of about $2.25\times 10^{36}\, \rm  erg\, s^{-1}$, which makes it among the top 20 pulsars as 
ranked by spin-down flux and thus a candidate to produce an X-ray emitting 
PWN \citep{Manchesteretal2005}. Recently, using an 
archival 30-ks {\it Chandra} observation of PSR~J1803-2137, \citet{CuiKonopelko2007} discovered
significantly extended X-ray emission around the pulsar (see also \citet{Kargaltsevetal2007}), 
indicating the presence of a PWN. The emission has a very hard spectrum, which is well 
described by a power law with photon index of about 1.2. The spatially-averaged flux is 
$\sim$$10^{-13}\, \rm erg\, cm^{-2}\, s^{-1}$ in the 0.3--10 keV band. The PWN is offset from 
the center-of-gravity of HESS~J1804-21 by about 10\arcmin. It is elongated, perpendicular to 
the pulsar's proper motion, 
suggesting that its X-ray emission probably emerges from a torus associated with the termination 
shock in the equatorial pulsar wind. Another example is HESS~J1809-193 and its plausible 
association with the pulsar PSR J1809-1917. \citet{Kargaltsev2007} has detected extended X-ray 
emission around the pulsar with {\it Chandra}, which is attributed to a PWN. As in the case of 
HESS~J1804-21, the PWN is also significantly offset from the center-of-gravity of HESS~J1809-193 
(by about 8\arcmin). 

There are several more HESS sources that also seem to be associated with pulsars but no 
corresponding PWNe have been seen yet. HESS~J1912+101 contains 
PSR~J1913+1011 \citep{aha2008b}, which is a rather old pulsar with a spin period of 36 ms 
and a spin-down age of $1.7 \times 10^5$~yrs, at a distance of 4.5 kpc. A radio pulsar was 
also detected in the Parkes Pulsar Survey, PSR~J1702-4128, near the tip of a tail-like 
extension of HESS~J1702-420 \citep{Krameretal2003}. It is a younger pulsar with a spin period 
of 182 ms and a spin-down age of about 55000 yrs and is located at a distance of about 5.2 kpc. 
Finally, PSR~J1617-5055 is located near the edge of HESS~J1616-508. It is an X-ray emitting, 
young pulsar with a spin period of 69~ms and a spin-down age of about 8000 yrs and is at 
a distance of about 6.5 kpc \citep{Toriietal1998}. 
In this work, we have undertaken a systematic search for extended X-ray emission associated with
PSR~J1702-4128, PSR~J1913+1011, and PSR~J1617-5055, using data from the archival {\it Chandra} 
observations, to take advantage of the superior spatial resolution of {\it Chandra}. We have 
analyzed all {\it Chandra} fields that contain the pulsars. The results from imaging and 
spectral analyses are reported here.

\section{Data Analysis and Results}

The {\it Chandra} data were extracted from the archival observations of RCW~103 (ObsID \#970), 
PSR~J1617-5055 (ObsID \#6684), PSR~J1702-4128 (ObsID \#4603), and PSR J1913+1011 (ObdID \#3854) 
with total exposure times of about 19~ks, 57~ks, 10.5~ks, and 19~ks, respectively. The data 
were all taken with the ACIS detector. Table~1 summarizes some of the key characteristics of 
these observations.

The data were reduced and analyzed with the standard {\it CIAO} analysis package (version 3.4), 
along with {\it CALDB 3.4.0}. For ObsID \#907, we followed the CIAO Science Threads\footnote{See 
http://asc.harvard.edu/ciao/threads/index.html} to prepare, filter, and reprocess the Level 1 
data to produce Level 2 data for subsequent analyses. For all other observations, we simply 
started with the Level 2 data that were derived from the archive, because they already 
incorporated the updated calibrations.

\subsection{Imaging Analysis}

We carried out a search for discrete sources in the 0.5--10~keV energy band with 
{\it celldetect}. This tool utilizes a sliding detection cell algorithm. The detection cell 
size was chosen to match the width of the local point spread function (PSF). Here we adopted
default values for all key parameters. For example, the size of the detection cell corresponds 
to the 80\% encircled energy area of the PSF and the signal-to-noise ratio threshold was set 
to 3. We excluded spurious detections in the vicinity of significant exposure variations, such 
as the detector edges or chip gaps etc. We estimated the statistical significance of each 
detection using the statistical method of \citet{LiMa1983}. For that we used the output of 
{\it celldetect}, which includes the sizes of the source and background regions, as well as 
the total number of counts extracted from the regions. Table~\ref{table2} summarizes 
the sources detected with a statistical significance greater than 4$\sigma$. 

\begin{figure}
\centering
\includegraphics[width=0.45\linewidth]{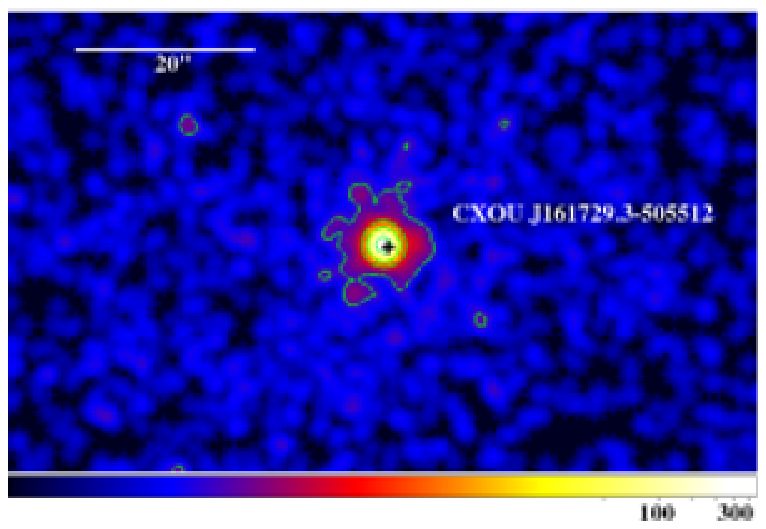} \\
\includegraphics[width=0.45\linewidth]{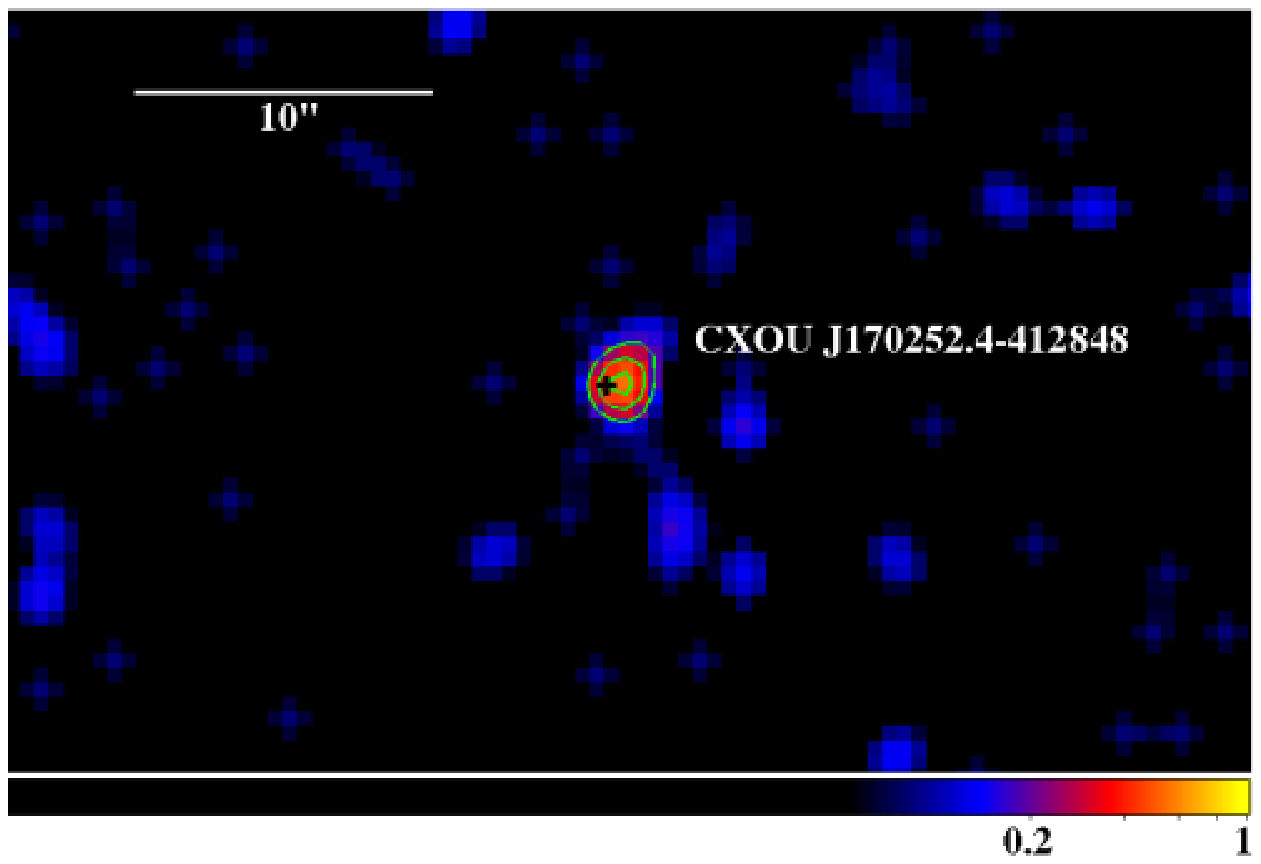} \\
\includegraphics[width=0.45\linewidth]{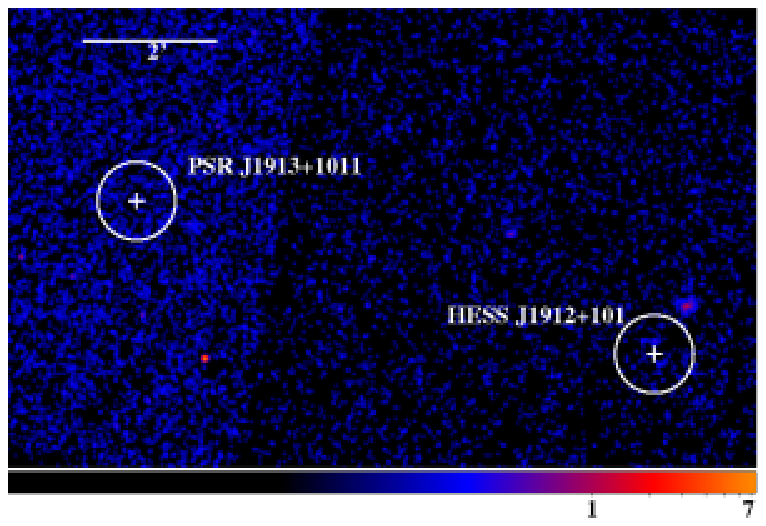}
\caption{{\it Chandra} X-ray images of the pulsar fields: PSR~J1617-5055 
(upper panel), PSR~J1702-4128 (middle panel), and PSR~J1913+1011 (low panel).
The crosses indicate the positions of the pulsars. All images were obtained 
from data in the 0.5--10~keV energy band. The contours correspond to the 
brightness levels of 0.4, 3, 25 and 200 counts (upper panel); 0.2, 0.35, and 
0.5 counts (middle panel). In the lower panel, the circles (in white) show the 
source regions used to derive flux upper limits. All images (in logarithmic 
scale) have been smoothed with a Gaussian kernel of 3 pixels in radius.
\label{fig2}}
\end{figure}

\begin{figure}
\psfig{figure=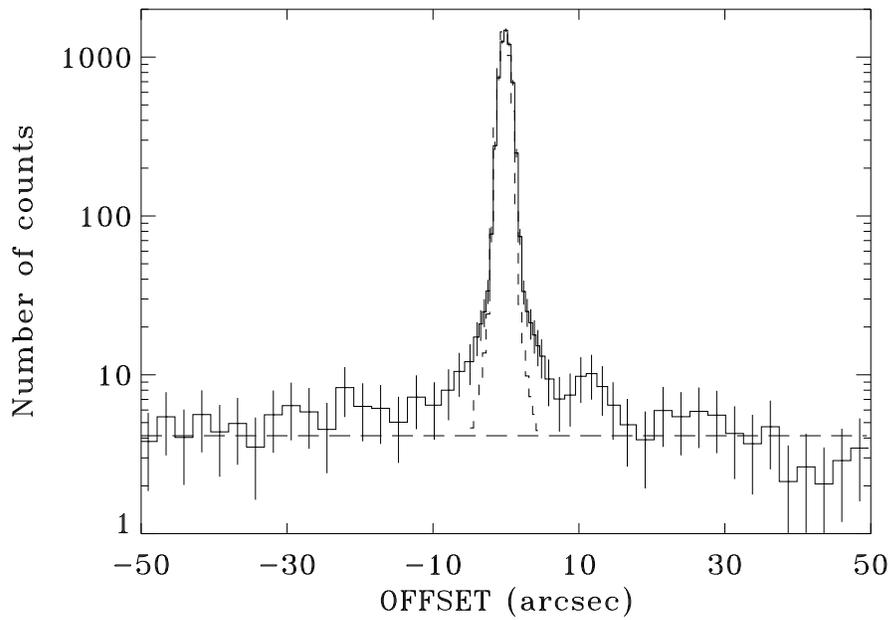,width=5in}
\caption{Linear profile of CXOU~J161729.3-505512 along right ascension (in solid histogram). 
It was made by projecting counts inside a horizontal $100\arcsec \times 30\arcsec$  
rectangular box that the source is centered on onto the axis of right ascension. Note that the 
data have been adaptively rebinned to obtain sufficient statistics. The background level is 
indicated by the long-dashed line. For comparison, the point spread function is overlaid 
(in dashed histogram).
\label{fig1}}
\end{figure}

For ObsID \#970, a total of ten sources were detected, including two sources on each of the I2, 
I3, and S4 chips and four sources on the S2 chip, respectively. CXOU~J161729.3-505512 coincides 
spatially with PSR~J1617-15055, which makes it a plausible counterpart of HESS~J1616-508. The 
source shows no apparent extension in this data set. However, a subsequent observation of 
longer exposure (ObsID \#6684) enabled us to resolve it as an extended X-ray source. To better 
identify the diffuse emission we ran {\it wavdetect} on this data set, and found that the 
effective $\sigma$ of the derived count distribution is about 2.36 times that of the PSF at the 
location of the source, indicating that the emission is significantly extended. 
Figure~\ref{fig2} shows an expanded view of the vicinity of CXOU~J161729.3-505512. From the 
raw count images, we constructed the radial profile of the source. Fitting the profile with a 
Gaussian function (plus a constant background) that is convolved with the local PSF resulted 
in a source extension of about 5.28$\pm$0.05 pixels (FWHM), which corresponds to about 
$2.6\arcsec$. However, the image also seems to show a more extended component of the emission.
To minimize the effects of asymmetry, we made a linear profile of the emission along right 
ascension, as shown in Figure~\ref{fig1}, with the PSF overlaid for comparison. We can clearly 
see that the emission extends up to roughly $20\arcsec$ in both directions. 

In ObsID \#4603, we detected one source on each of the I2 and S3 chips but observed no apparent
X-ray emission around the position of PSR~J1702-4128. Only after we lowered the signal-to-noise 
threshold used by {\it celldetect} down to 1.5, an X-ray source (CXOU~J170252.4-412848) emerged. 
To evaluate the statistical significance of the detection, we extracted source counts from a 
circular region of 7 pixels in radius and background counts from another circular region of 20 
pixels in radius. Based on the Li \& Ma method, we derived a significance of about 4.6$\sigma$. 
We also ran {\it wavdetect} on this {\it Chandra} field and found that the 
effective $\sigma$ of the count distribution is about 1.46 times that of the PSF at the location 
of the source, suggesting that CXOU~J170252.4-412848 might also be a slightly extended source.
There are too few counts to meaningfully quantify the extension. The image is also shown in 
Figure~\ref{fig2}. The source is located almost exactly at the position of the pulsar, which 
makes it a plausible X-ray counterpart of the latter. 

For ObsID \#3854, a few sources have been detected one of them on the I2 chip, three others
on each of the S2 and S3 chips, and two more on the S4 chip. No apparent X-ray emission can be 
seen around the position of PSR J1913+1011 nor near the center-of-gravity of HESS J1912+101. 
We also carried out more sensitive searches, with {\it wavdetect}, for faint extended emission 
at both locations but failed to detect any. To derive flux upper limits, we used a circular 
source region of a 35$\arcsec$ radius around each of the positions and a similar circular 
background region nearby. The measured count rates were converted into the corresponding fluxes 
by adopting the line-of-sight hydrogen column densities \citep{Dickey1990} and assuming a 
power-law spectrum of photon index 2. The results are summarized in Table~\ref{table3}.

We searched through the SIMBAD and NED databases for the counterparts of all detected sources 
within an angular offset of less than $30\arcsec$ from each source. Besides the pulsars of 
interest, two of the sources, CXOU~J161723.7-505150 and CXOU~J191316.1+100902, appear to be 
associated with the stars, HD~146184 and HD~179712, respectively. In addition, two other 
sources, CXOU~J161727.9-505549 and CXOU~J191338.4+101200, are about $24\arcsec$ and $25\arcsec$ 
away from IRAS~16137-5048 and IRAS~19112+1007, respectively.

\subsection{Spectral Analysis}

The CIAO tool {\it specextract} was used to extract the X-ray spectrum of CXOU~J161729.3-505512, 
which is possibly a PWN associated with the pulsar PSR~J1617-5055. We used a circular region of
a 50 pixel radius ($\sim$25$\arcsec$) centered on the best-fit position of the source 
to extract source counts and a concentric annulus with an inner radius of 60 pixels and an 
outer radius of 100 pixels to extract background counts. This tool enabled to generate both the 
overall (source+background) and background spectra, as well as the corresponding response matrix 
files and auxiliary response files for subsequent spectral modeling. 

\begin{figure}
\psfig{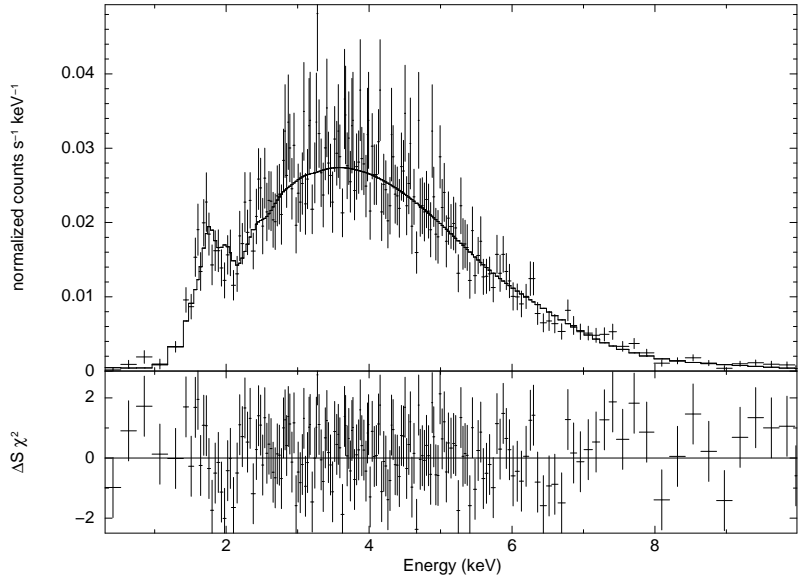}
\caption{X-ray spectrum of CXOU~J161729.3-505512. The solid histogram shows 
the best power-law fit to the data. The residuals of the fit are shown in 
the bottom panel. }
\label{fig3}
\end{figure}

We modeled the spectra with {\it XSPEC} \citep{Arnaud1996}. The data points below 0.3~keV 
or above 10~keV were excluded an the remaining data rebinned to achieve at least 15 counts in 
each energy bin. The obtained spectrum can be fitted very well with an absorbed power-law model, 
as shown in Figure~\ref{fig3}, with a reduced $\chi^2$ about 0.94 for 660 degrees of freedom.
The best-fit parameters are shown in Table~\ref{table3}. We should note that the pile-up
effects are quite small ($\sim$3\%), thanks to the 1/4--subarray readout mode adopted.

We attempted the same analyses for CXOU~J170252.4-412848 (but with the source and background 
regions used in the imaging analysis). Only about 11 net source counts 
were obtained, which is too few to warrant any reliable spectral modeling. Purely for the
purpose of estimating the flux of the source, we fitted the data to an absorbed  
power-law spectrum with the photon index fixed at $\Gamma=2$. The flux is  
$(0.2^{+0.2}_{-0.1}) \times 10^{-13}$ erg cm$^{-2}$ s$^{-1}$ in 0.3 -- 10 keV energy band. These 
results are also shown in Table~\ref{table3}.

\section{Discussion}

One of the most significant recent developments in high energy astrophysics is the detection
of a wide variety of $\gamma- $ray sources at the very high-energy energies (above 100~GeV; 
see, e.g., \citet{Cui2006}, for a review). Identifying counterparts of these sources at other 
longer wavelengths can help unveiling the physical mechanism responsible for the VHE 
$\gamma- $ray emission. Among the Galactic VHE $\gamma- $ray sources PWNe represent the most
populous class. In many cases the association of a VHE $\gamma- $ray source with a PWN is
established by combining the positional and morphological similarities observed in various 
wavelength ranges. Further support comes from the successful modeling of their broadband 
multi-wavelength energy spectra. In some cases the evidence for an association is less 
compelling or non-existent and thus requires further observations. 
A strong argument in favor of plausible association of a PWN with the a VHE $\gamma- $ray 
source is the relatively low integral efficiency needed to convert some of the pulsar's 
spin-down luminosity to VHE $\gamma- $rays (e.g., see, \citet{Gallant2006}). If one defines 
the efficiency as $\epsilon = (4 \pi d^2 F_{\gamma})/\dot{E}$, where $F_\gamma$ is the 
integral VHE $\gamma- $ray flux above 300~GeV measured with the HESS instrument, it ranges
from roughly 1\% to 10\% for most of the known TeV sources that are associated with PWNe. 
Two caveats should be pointed out. First of all, the pulsar distance is often poorly 
determined. Secondly, this approach largely elides any substantial change in pulsar's 
spin-down luminosity during early evolution of the pulsar and its PWN.                 

In this work, we used archival {\it Chandra} data to search for X-ray counterparts of three 
unidentified TeV $\gamma- $ray sources that appear to be associated with known radio pulsars. 
We have likely detected an X-ray emitting PWN associated with PSR~J1617-5055, which could 
also be the underlying engine of HESS~J1616-508. Figure~3 shows all the X-ray sources 
detected in the vicinity of HESS~J1616-508. Among them only one, CXOU~J161729.3-505512, 
is spatially extended. This, coupled with its spatial coincidence with PSR~J1617-5055, 
is strongly in favor of a PWN origin of the X-ray emission, which also makes it a promising
candidate for the X-ray counterpart of HESS~J1616-508. \citet{Gallant2006} has estimated the 
integrated $\gamma- $ray flux in the 0.3--30 TeV band, about 
$3.7 \times 10^{-11} \rm \, erg \, cm^{-2} \, s^{-1}$ or about 1.3\% of the spin-down flux 
of the pulsar, which is quite achievable. We should note that the X-ray emission associated
with PSR~J1617-5055 has been seen before, most clearly with the {\it XMM-Newton} 
data \citep{Becker2002,Landietal2007}. However, for the first time, we have spatially 
resolved the emission, thanks to the superior resolution of {\it Chandra}. Combining data from 
{\it XMM-Newton}, {\it INTEGRAL}, and {\it BeppoSAX}, \citet{Landietal2007} derived the
following spectral parameters from a simple power-law fit: 
$N_H = 3.87^{+0.36}_{-0.28} \times 10^{22}$ $cm^{-2}$, $\Gamma = 1.42^{+0.12}_{-0.10}$,
and the unabsorbed 2-10 keV flux $F_X = 4.2 \times 10^{-12}\rm  \, erg\, cm^{-2} \, s^{-1}$. 
Their spectrum thus seems a bit softer than the one that we derived here from the 
{\it Chandra} observation, although the hydrogen column density and flux are in general 
agreement (the unabsorbed 2-10 keV flux is about 
$4.0 \times 10^{-12}\rm  \, erg\, cm^{-2} \, s^{-1}$ in our case). A possible explanation
for the discrepancy is that they might have included more diffuse emission in the surrounding
region, given the lower spatial resolution of the instruments used (especially {\it INTEGRAL} 
and {\it BeppoSAX}). \citet{Becker2002} found that about 53\% of the X-ray flux is pulsed and 
thus associated with the pulsar. Therefore, the PWN contribution is no more than 47\% of the
measured flux (or $1.6 \times 10^{-12}\rm  \, erg\, cm^{-2} \, s^{-1}$ in the 0.3--10 keV band). 
A lower limit on the PWN contribution can be obtained from the fraction of the X-ray emission 
outside the PSF. From Fig.~\ref{fig1}, we estimated that at least $\sim$10\% of the observed 
flux can be attributed to the PWN.

We also see evidence for X-ray emission associated with PSR~J1702-4128. The emission might be 
slightly extended. There are also other sources detected in the vicinity of HESS~J1702-420, as 
shown in Fig.~4, but the pulsar/PWN connection makes CXOU~J170252.4-412848 a more likely 
counterpart of the X-ray source. If HESS~J1702-420 is powered by the pulsar, its measured
VHE $\gamma- $ray flux ($1.4 \times 10^{-11}\rm \, erg\, cm^{-2} \, s^{-1}$) would correspond
to about 11\% of the spin-down flux of the pulsar \citep{Gallant2006}, which is rather high 
but not implausible. We failed to detect any X-ray emission at either the position of 
PSR~J1913+1011 or the center-of-gravity (CoG) of HESS~J1912+101 (although CXOU J191247.0+100948 
is only about 52\arcsec\ from the CoG), down to a flux level of roughly
$1 \times 10^{-13}$ $erg$ $cm^{-2}$ $s^{-1}$ or lower (see Table~3). If HESS~J1912+101 is
powered by the pulsar, the measured VHE $\gamma- $ray flux accounts for only about 1\% of the
spin-down flux of the pulsar \citep{aha2008b}. The non-detection of X-ray emission 
associated with the 
pulsar does not necesarily contradict a possible association between the TeV $\gamma$-ray 
source and the pulsar. For instance, different populations of relativistic electrons could be 
responsible for the synchrotron X-ray emission and the TeV $\gamma$ rays seen from this source
(see, e.g. \citet{Funk2007}). 

\section*{Acknowledgments}

We thank the anonymous referee for bringing to our attention the long observation of 
PSR~J1617-5055 (ObsID \#6684) and for many detailed and constructive comments, which have 
helped improve the manuscript significantly. 
This research has made use of the NASA/IPAC Extragalactic Database (NED), which is operated by 
the Jet Propulsion Laboratory, California Institute of Technology, under contract with the 
National Aeronautics and Space Administration (NASA), and of the SIMBAD Database. We gratefully 
acknowledge financial support from the Department of Energy and NASA.

\begin{deluxetable}{ccccc}
\tablecolumns{5}
\tablewidth{0pt}
\tablecaption{Observations.}
\tablehead{
\colhead{ObsID}&\colhead{Pulsar}&\colhead{Exposure}&\colhead{Offset\tablenotemark{a}}&\colhead{ACIS Chips}\\
\colhead{}&\colhead{}&\colhead{Time(ks)}&\colhead{($\arcmin$)}&\colhead{}}
\startdata
\#970&PSR J1617-5055&19&6.36& I2, I3, S2, S3, S4\\
\#6684&PSR J1617-5055&57&0.80& I3\\
\#4603&PSR J1702-4128&10.5&0.67& I0, I1, I2, I3, S2, S3\\
\#3854&PSR J1913+1011&19&0.60& I2, I3, S1, S2, S3, S4\\
\enddata
\tablenotetext{a}{From the aim point.}
\end{deluxetable}

\begin{table}
\caption{Detected X-ray Sources$^{\dag}$ \label{table2}}
\setlength{\tabcolsep}{1mm}
\begin{tabular}{cllll}\hline\hline
ObsID&Source&Right ascension&Declination&Count rate\\
& &J2000&J2000&10$^{-3}$ cts s$^{-1}$\\\hline
&CXOU J161723.7-505150&16:17:23.73(4)&-50:51:50.3(4)&4.5$\pm$0.6\\
&CXOU J161727.9-505549&16:17:27.970(3)&-50:55:49.7(4)&1.5$\pm$0.1\\
&CXOU J161729.3-505512&16:17:29.353(1)&-50:55:12.77(1)&69$\pm$1\\
&CXOU J161734.3-511227&16:17:34.3(1)&-51:12:27.8(8)&1.6$\pm$0.4\\
&CXOU J161741.9-511006&16:17:41.90(4)&-51:10:06.3(4)&2.6$\pm$0.4\\
\raisebox{1.5ex}[0cm][0cm]{970/6648}&CXOU J161747.2-505709&16:17:47.28(2)&-50:57:09.1(2)&0.9$\pm$0.2\\
&CXOU J161820.7-510737&16:18:20.79(3)&-51:07:37.1(2)&7.4$\pm$0.6\\
&CXOU J161844.0-505728&16:18:44.1(1)&-50:57:29(1)&1.2$\pm$0.3\\
&CXOU J161849.2-510424&16:18:49.29(8)&-51:04:24.2(7)&2.4$\pm$0.5\\
&CXOU J161908.4-505507&16:19:08.5(1)&-50:55:08(1)&1.9$\pm$0.4\\ \hline
&CXOU J170238.5-413311&17:02:38.52(2)&-41:33:11.6(2)&4.5$\pm$0.6\\
4603&CXOU J170252.4-412848&17:02:52.48(1)&-41:28:48.2(1)&1.1$\pm$0.2\\
&CXOU J170333.5-413055&17:03:33.55(5)&-41:30:55.2(6)&1.7$\pm$0.4\\ \hline
&CXOU J191238.0+101043&19:12:38.01(2)&+10:10:43.4(3)&8.9$\pm$0.8\\
&CXOU J191240.6+101755&19:12:40.63(4)&+10:17:55.2(7)&2.2$\pm$0.4\\
&CXOU J191245.2+100656&19:12:45.28(6)&+10:06:56.9(7)&1.3$\pm$0.3\\
&CXOU J191247.0+100948&19:12:47.00(3)&+10:09:48.4(4)&1.9$\pm$0.4\\
3854&CXOU J191316.1+100902&19:13:16.192(4)&+10:09:02.06(4)&4.6$\pm$0.5\\
&CXOU J191331.9+101231&19:13:31.93(1)&+10:12:31.5(2)&1.2$\pm$0.2\\
&CXOU J191338.4+101200&19:13:38.41(2)&+10:12:00.2(3)&1.2$\pm$0.3\\
&CXOU J191351.1+101152&19:13:51.16(3)&+10:11:52.1(5)&1.9$\pm$0.4\\
&CXOU J191400.8+101403&19:14:00.88(7)&+10:14:03.5(9)&1.5$\pm$0.3\\
\hline 
\end{tabular}

\vspace*{2mm}
$^{\dag}$ The numbers in parentheses indicate uncertainty in the last digit. Note that only 
statistical uncertainties are shown. \\
\end{table}

\begin{table}
\caption{Properties of X-ray Emission$^{\dag}$ \label{table3}}
\setlength{\tabcolsep}{1mm}
\begin{tabular}{lccc}\hline\hline
Name&N$_{\rm H}$&$\Gamma$&Flux (0.3--10 keV) \\
&(10$^{22}$ cm$^{-2})$&&$(10^{-13} \, \rm erg\, cm^{-2} \, s^{-1}$) \\ \hline
PSR J1617-5055&3.3$\pm$0.3&1.1$\pm$0.1&34$^{+4}_{-7}$ \\
PSR J1702-4128&1.8$^{+0.6}_{-0.5}$&2.0 (fixed) &0.2$^{+0.2}_{-0.1}$ \\
PSR J1913+1011&1.79$^c$&2.0 (fixed) &$<$0.74$^b$ \\
HESS CoG$^a$&1.81$^c$ &2.0 (fixed) &$<$0.7$^b$ \\
\hline\hline
\end{tabular}

\vspace*{2mm}
$^{\dag}$ The columns are: source name, hydrogen column density, photon index, and flux. The 
errors shown represent 90\% confidence intervals, unless otherwise noted. \\ 
$^a$ The center of gravity of HESS~J1912+101 (see Figure~\ref{fig2}). \\
$^b$ $3\sigma$ upper limits. \\
$^c$ Taken from \citet {Dickey1990}\\
\end{table}

\clearpage
\pagebreak

\begin{figure}
\psfig{figure=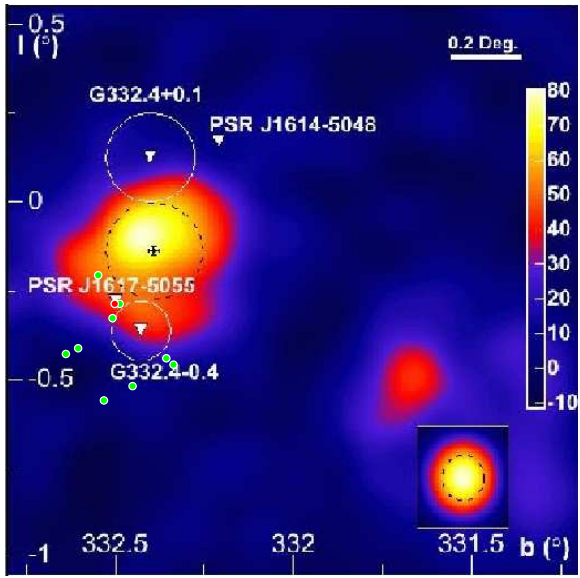,width=3in}
\caption{X-ray sources in the vicinity of HESS~J1616-508. The TeV 
$\gamma- $ray image was adapted from \citet{aha2006b}. The positions of the 
detected X-ray sources are indicated by green filled circles. The red filled 
circle indicates the position of CXOU~J161729.3-505512, 
which is likely associated with PSR~J1617-5055 and its PWN. \label{fig4}}
\end{figure}

\begin{figure}
\psfig{figure=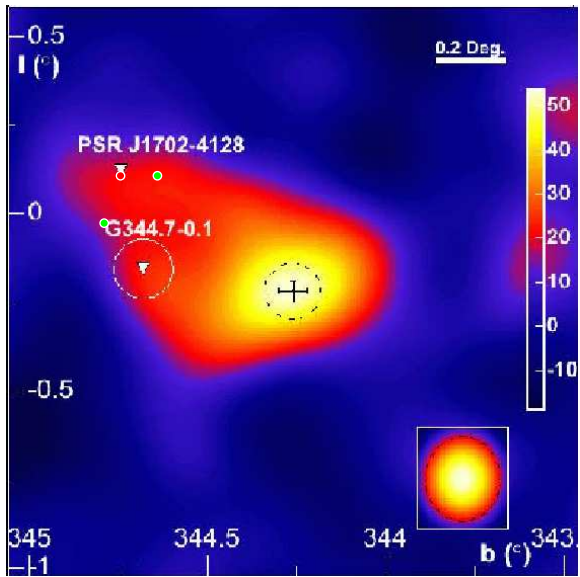,width=3in}
\caption{Same as Fig.~3, but for HESS~J1702-420. The red filled circle shows 
the position of CXOU~170252.4-412848, which might be associated with 
PSR~J1702-4128 and its PWN. \label{fig5}} 
\end{figure}

\end{document}